\documentclass[conference]{IEEEtran}
\IEEEoverridecommandlockouts
\usepackage{cite}
\usepackage{amsmath,amssymb,amsfonts}
\usepackage{algorithmic}
\usepackage{graphicx}
\usepackage{booktabs}
\usepackage{hyperref}
\usepackage{textcomp}
\usepackage{xcolor}
\usepackage{subcaption}
\usepackage{xcolor} % for color
\def\BibTeX{{\rm B\kern-.05em{\sc i\kern-.025em b}\kern-.08em
    T\kern-.1667em\lower.7ex\hbox{E}\kern-.125emX}}

\def\BibTeX{{\rm B\kern-.05em{\sc i\kern-.025em b}\kern-.08em
    T\kern-.1667em\lower.7ex\hbox{E}\kern-.125emX}}

\makeatletter
\newcommand{\linebreakand}{%
  \end{@IEEEauthorhalign}
  \hfill\mbox{}\par
  \mbox{}\hfill\begin{@IEEEauthorhalign}
}
\makeatother

\begin{document}

% Create a separate title page for the copyright statement
% \begin{titlepage}
% \thispagestyle{empty} % No headers or footers on this page
% \vspace*{\fill} % Vertically center the content on the page
% \begin{center}
%     \textbf{Copyright Statement}\\[1cm]
%     This paper has been accepted for publication in the IEEE ICBASE 2024 conference \\

%     © 2024 IEEE. Personal use of this material is permitted. Permission from IEEE must be obtained for all other uses, in any current or future media, including reprinting/republishing this material for advertising or promotional purposes, creating new collective works, for resale or redistribution to servers or lists, or reuse of any copyrighted component of this work in other works.
% \end{center}
% \vspace*{\fill}
% \end{titlepage}

\title{Identification of Prognostic Biomarkers for Stage III Non-Small Cell Lung Carcinoma in Female Nonsmokers Using Machine Learning}

\author{
    \IEEEauthorblockN{Huili Zheng\textsuperscript{1a,*}, Qimin Zhang\textsuperscript{1b}, Yiru Gong\textsuperscript{2a}, Zheyan Liu\textsuperscript{2b}, and Shaohan Chen\textsuperscript{2c}}
    \IEEEauthorblockA{
        \textsuperscript{1, 2}\textit{Department of Biostatistics, Columbia University, New York, NY 10032, USA}
    }
    \IEEEauthorblockA{
        hz2710@caa.columbia.edu\textsuperscript{1a,*}, qimin.zhang@columbia.edu\textsuperscript{1b},\\
        yiru.g@columbia.edu\textsuperscript{2a}, zl3119@caa.columbia.edu\textsuperscript{2b}, shaohan.chen@caa.columbia.edu\textsuperscript{2c}
    }
}

\maketitle

\begin{abstract}
Lung cancer remains a leading cause of cancer-related deaths globally, with non-small cell lung cancer (NSCLC) being the most common subtype. This study aimed to identify key biomarkers associated with stage III NSCLC in non-smoking females using gene expression profiling from the GDS3837 dataset. Utilizing XGBoost, a machine learning algorithm, the analysis achieved a strong predictive performance with an AUC score of 0.835. The top biomarkers identified—CCAAT enhancer binding protein alpha (C/EBP$\alpha$), lactate dehydrogenase A4 (LDHA), UNC-45 myosin chaperone B (UNC-45B), checkpoint kinase 1 (CHK1), and hypoxia-inducible factor 1 subunit alpha (HIF-1$\alpha$)—have been validated in the literature as being significantly linked to lung cancer. These findings highlight the potential of these biomarkers for early diagnosis and personalized therapy, emphasizing the value of integrating machine learning with molecular profiling in cancer research.
\end{abstract}


\section{Introduction}

Lung cancer is a significant health challenge globally, being among the most common and deadly cancers. It accounts for approximately 12\% of all new cancer diagnoses and nearly 20\% of all cancer deaths annually. It is primarily classified into two major types: non-small cell lung cancer (NSCLC), which represents about 82\% of cases, and small cell lung cancer (SCLC), which is less common but more aggressive. Despite advancements in early detection and treatment, the prognosis for lung cancer patients remains poor, with a five-year survival rate of approximately 25\%, underscoring the critical need for continued research and innovation in this field \cite{Siegel2024CancerStatistics}.

Smoking is widely recognized as the leading risk factor for lung cancer, responsible for approximately 80\% to 90\% of all lung cancer cases. The carcinogenic compounds in tobacco smoke contribute significantly to the development of both NSCLC and SCLC. However, an increasing number of lung cancer cases are being diagnosed in individuals who have never smoked, particularly among women. This trend suggests that other factors, like genetic predispositions, environmental exposures (e.g., radon and air pollution), and hormonal influences, may play a critical role in lung cancer development among non-smokers. To enhance survival rates in non-smoking lung cancer patients, it is crucial to conduct a thorough analysis of the molecular mechanisms underlying carcinogenesis in NSCLC. This approach will help identify more effective biomarkers for early diagnosis and uncover new targets for drug development. \cite{Siegel2024CancerStatistics, Sun2023NonSmokingLungCancer}.

Stage III non-small cell lung cancer is a complex disease, where accurate prognostic evaluation is key to personalized treatment. Detecting prognostic biomarkers can guide therapeutic decisions, potentially benefiting patients eligible for immunotherapy or targeted therapies. Recent studies, including a review on stage III NSCLC management, highlight the potential of biomarker-driven approaches to refine patient selection, improve outcomes, and advance personalized treatment strategies in this challenging lung cancer subset \cite{Kim2024StageIIINSCLC}. Given the poor prognosis of stage III NSCLC, reliable biomarkers could significantly enhance treatment effectiveness and survival rates \cite{Malapelle2021EmergingBiomarkers}.

Gene expression profiling measures the activity of thousands of genes simultaneously, offering insights into the molecular mechanisms of diseases like cancer. RNA is extracted from tissue samples, converted to cDNA, and analyzed using high-throughput technologies such as microarrays or next-generation sequencing (NGS). This data identifies differentially expressed genes, which can serve as biomarkers for diagnosis, prognosis, and therapeutic targeting \cite{Wang2023GeneExpressionTechniques, Smith2022NGSApplications}.

The rapid expansion of the bioinformatics field such as gene expression profiling has also led to a growing dependence on machine learning techniques\cite{Weng2024bigdata} for diagnosing and predicting complicated diseases with their biomarkers. However, the biomedical data is often high-dimensional, with numerous variables and limited data points, along with the issue of multicollinearity, which often complicates analysis. To tackle these challenges, researchers have created a range of machine learning algorithms specifically for bioinformatics, such as tree-based methods\cite{shen2024harnessing} as well as neural networks, which have been proven powerful for identifying complex patterns for different domains, such as computer vision\cite{zhang2024cunetunetarchitectureefficient, xin2024mmap}, natural language processing, and more comprehensive tasks\cite{xin2024vmt, peng2024lingcn, liu2024applicationofmultimodel}.

\section{Data}
This study utilized the publicly available dataset GDS3837 \cite{Lu2010SEMA5A} from the National Center for Biotechnology Information (NCBI) Gene Expression Omnibus (GEO) database, which provides comprehensive gene expression data for lung tissue specimens. The dataset includes expression profiles derived from both tumor and adjacent normal lung tissue, offering valuable insights into the molecular differences associated with lung cancer, particularly in non-smoking individuals. By leveraging this dataset, the study aimed to identify potential biomarkers and molecular signatures that could contribute to the understanding of lung cancer pathogenesis and aid in the development of targeted therapies.

\subsection{Data Collection}
60 pairs of tumor and adjacent normal lung tissue specimens were collected from non-smoking females at National Taiwan University Hospital and Taichung Veterans General Hospital, with written informed consent obtained from all participants. The lung tissues were promptly preserved in RNAlater buffer, quickly snap-frozen in liquid nitrogen, and stored at \textminus 80°C for subsequent RNA extraction. Of these, 60 sample pairs passed quality control and were processed for gene expression profiling. Total RNA was isolated using TRIzol reagent and purified with the RNeasy mini kit. Double-strand cDNA and cRNA were synthesized from purified RNA and hybridized to GeneChip Human Genome U133 Plus 2.0 expression arrays (Affymetrix). After 16 hours of hybridization, the arrays were washed, scanned, and the resulting data were analyzed for mRNA expression levels using Partek software. The analysis included background correction, quantile normalization, and summarization through robust multiarray average analysis. \cite{Lu2010SEMA5A, Lu2015SNPs}.

\subsection{Data Characteristic}
\begin{figure}[hbt!]
    \centering
    \includegraphics[width=0.9\linewidth]{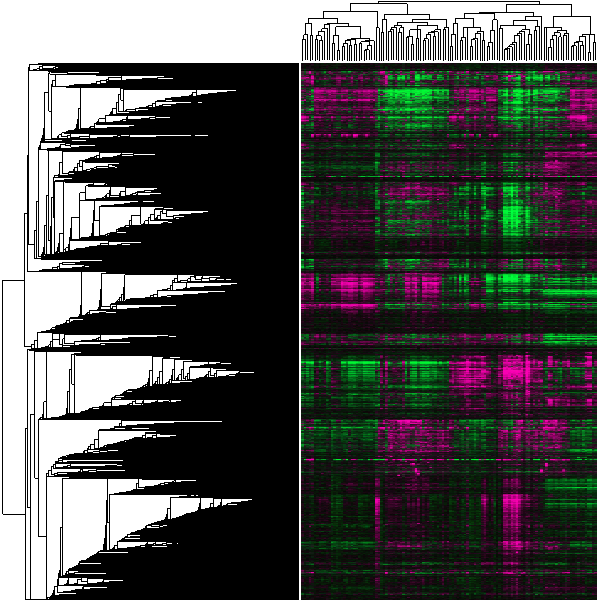}
    \caption{GDS3837 Gene Expression Heatmap}
    \label{fig:GDS3837}
\end{figure}

The heatmap Fig.~\ref{fig:GDS3837} represents gene expression data from the GDS3837 dataset. It displays expression levels of 54,675 genes across 120 samples. The hierarchical clustering reveals patterns of co-expression, but the dense clustering indicates significant multicollinearity in the data, which complicates the identification of independent biomarkers and requires advanced analytical methods to address. 

\begin{table}[h!]
\centering
\caption{Clinical Characteristics of the Patients}
\begin{tabular}{@{}l@{\hspace{0.3cm}}l@{\hspace{0.5cm}}c@{\hspace{0.5cm}}c@{}}  % Adjust hspace for spacing
\hline
\textbf{Characteristics} & & \textbf{Sample size, n (\%)} & \textbf{Age (mean $\pm$ SD, y)} \\
\hline
\textbf{Sex} & & & \\
 & Female & 60 (100) & 61 $\pm$ 10 \\
\hline
\textbf{Tumor stage} & & & \\
 & I + II & 47 (78) & 61 $\pm$ 11 \\
 & III & 13 (22) & 61 $\pm$ 7 \\
\hline
\end{tabular}
\end{table}

\subsection{Data Pre-Processing and Split}

In this study, we did a train-test split on 60 individuals, with 60\% (n=36) used for training and 40\% (n=24) for testing. This approach ensures that the model has sufficient data for learning while retaining enough for evaluating its performance on unseen individuals.

For the target variable, we selected patients with stage III or later lung cancer as positive class and others as negative class.

\section{Methods}
XGBoost\cite{xgboost}, a tree-based machine learning method, was utilized in this study to identify biomarkers associated with stage III lung cancer. The objective function in XGBoost is designed to minimize the following expression: 

\begin{equation} 
\mathcal{L}(T) = \sum_{i=1}^{n} l(y_i, \hat{y}i) + \lambda \sum{t=1}^{|T|} \Omega(f_t) 
\label{eq} 
\end{equation} 
In this equation, $T$ is the collection of tree models, $l(y, \hat{y})$ is a convex loss function that's differentiable, $y_i$ and $\hat{y}_i$ represents the true output and the predicted output for the $i$-th data point respectively, and $\Omega(f_t)$ is the penalty term\cite{jin2024learning} for each tree model $f_t$ in the collection.

XGBoost optimizes\cite{chen2021pareto} the loss function described in Equation~\ref{eq} through an additive approach, where a new tree is sequentially added to the model to improve the overall prediction. This iterative process can be expressed as: 

\begin{equation} 
\hat{y}^{(p)}(x) = \hat{y}^{(p-1)}(x) + \eta f_p(x) 
\end{equation} 

where $\hat{y}^{(p)}(x)$ is the prediction after $p$ iterations, $\hat{y}^{(p-1)}(x)$ is the prediction from the previous iteration, $f_p(x)$ represents the newly added tree in the $p$-th iteration, and $\eta$ is the learning rate.

By showcasing the potential of and XGBoost, the research\cite{zheng2024advancedpayment} of Yu et al. lays a solid groundwork and guidance for future research aimed at developing robust and efficient classification models for a range of applications. We have leveraged Chang's research findings\cite{yu2024advanced}, significantly enhancing our model processing efficiency through the adoption of his data handling techniques. As a tree-based model, XGBoost also has a good explainability\cite{tao2019thefact}, so we anticipate that it will capture complex patterns within the GDS3837 dataset to identify biomarkers associated with stage III lung cancer.

To evaluate the model's performance, we used the Receiver Operating Characteristic (ROC) score, which measures the model's ability to distinguish between classes.

\section{Results}
\begin{figure}[hbt!]
    \centering
    \includegraphics[width=1\linewidth]{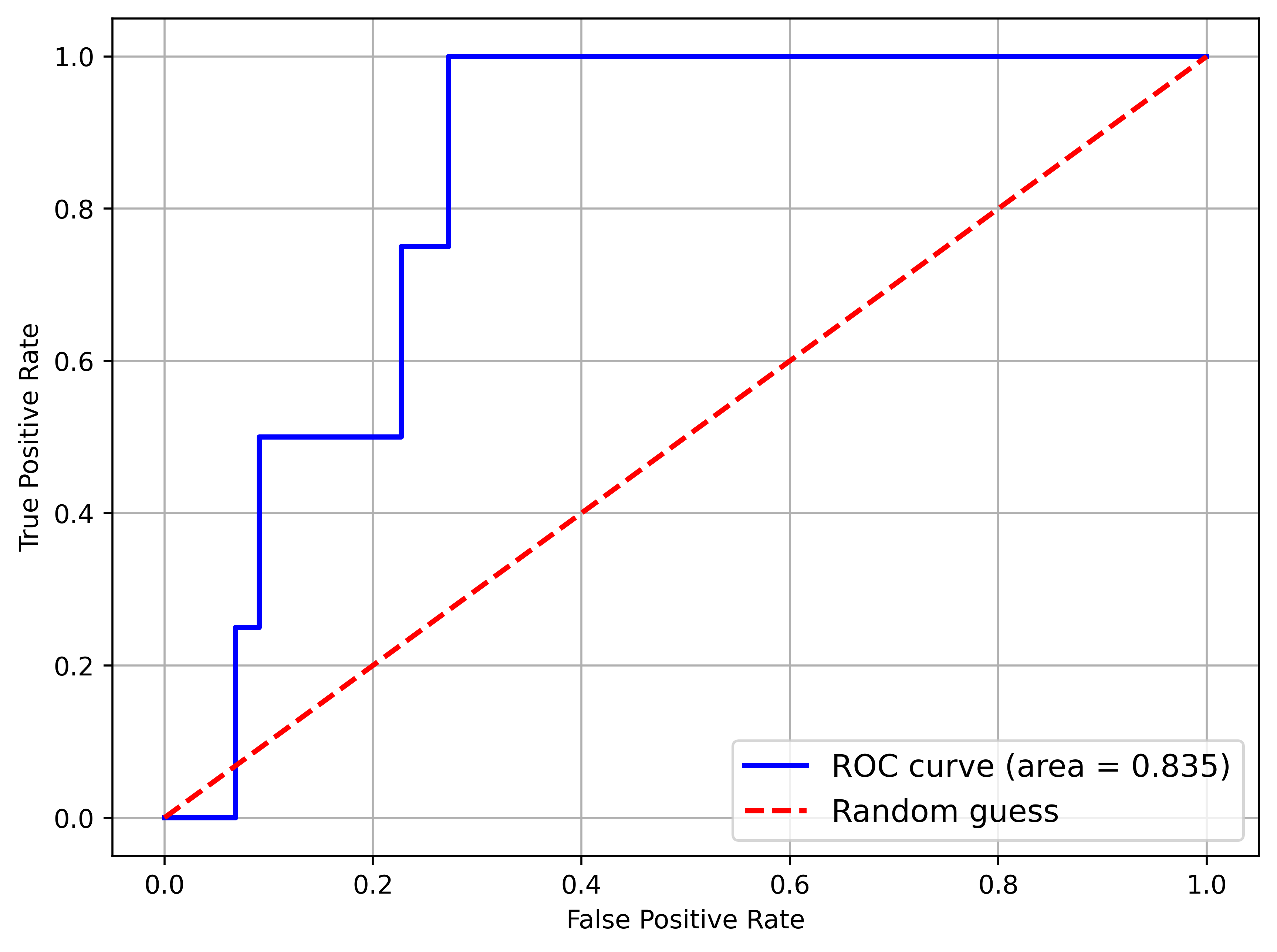}
    \caption{ROC Curve of the Model}
    \label{fig:roc_curve}
\end{figure}

As illustrated in this Matplotlib-visualized\cite{Weng2024fortifying} Fig. \ref{fig:roc_curve}, XGBoost demonstrated outstanding performance in predicting stage III lung cancer using gene expression profiling data, achieving an impressive AUC score of 0.835. This high AUC score reflects the model's strong ability to distinguish between patients with and without stage III lung cancer, underscoring the effectiveness of XGBoost in capturing complex patterns within high-dimensional genomic data. The model's robustness and accuracy make it a powerful tool for identifying critical biomarkers associated with stage III lung cancer, offering significant potential for improving early detection and personalized treatment strategies.

\begin{figure}[hbt!]
    \centering
    \includegraphics[width=1\linewidth]{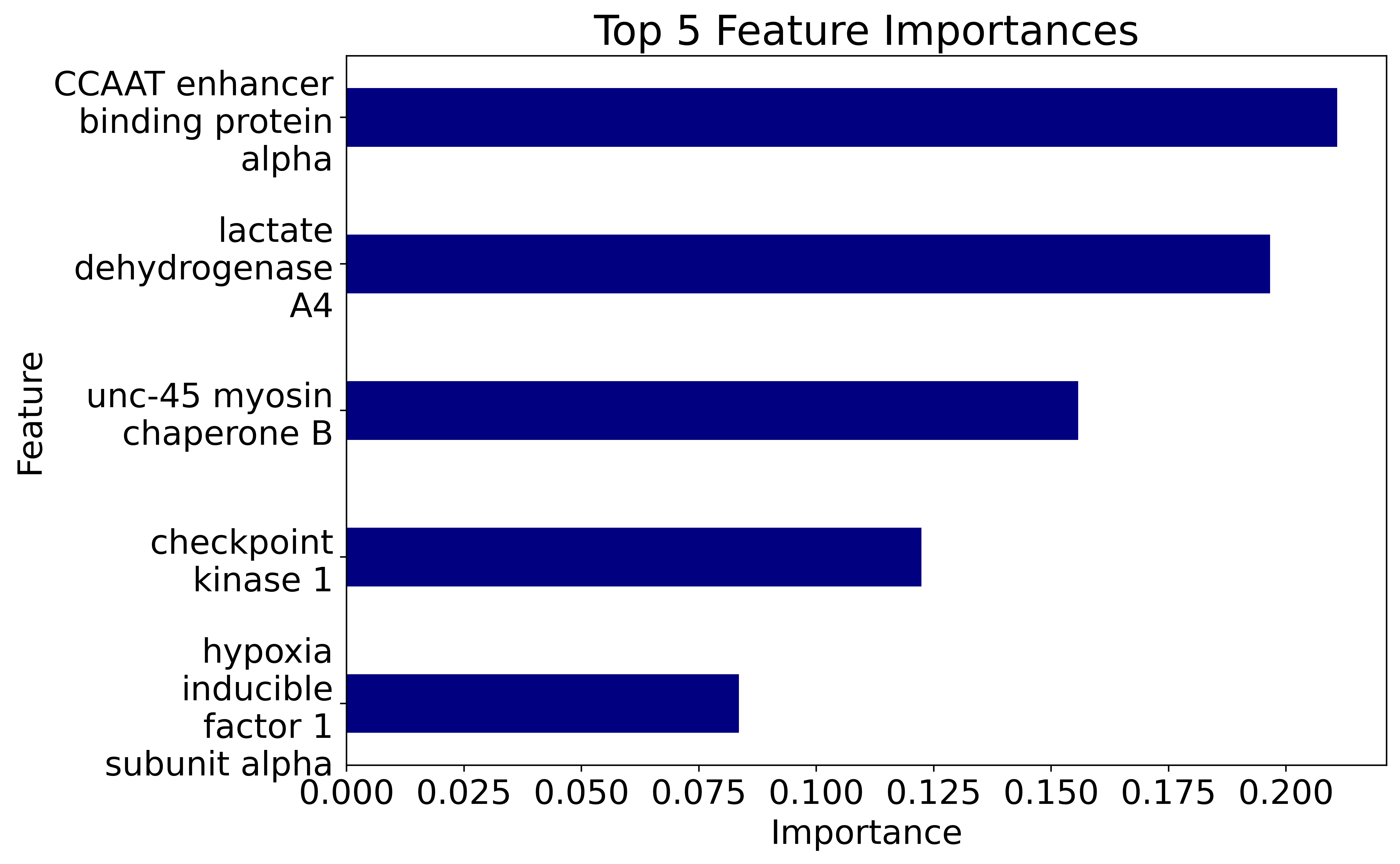}
    \caption{Feature Importance}
    \label{fig:feature_importance}
\end{figure}

As displayed in the feature importance Fig. \ref{fig:feature_importance}, the top 5 biomarkers selected by the XGBoost model in this study include critical genes and proteins associated with various cellular functions and responses. These biomarkers are: CCAAT enhancer binding protein alpha (C/EBP$\alpha$), a key regulator of cellular differentiation and proliferation; lactate dehydrogenase A4 (LDHA), an enzyme involved in glycolysis and often linked to cancer metabolism; unc-45 myosin chaperone B (UNC-45B), which plays a role in muscle cell organization and function; checkpoint kinase 1 (CHK1), a vital component of the DNA damage response and cell cycle control; and hypoxia inducible factor 1 subunit alpha (HIF-1$\alpha$), a transcription factor crucial for cellular adaptation to low oxygen conditions. The selection of these features by XGBoost underscores their potential importance as biomarkers in the context of lung cancer, particularly in non-smoking females, and highlights their relevance in pathways critical to cancer progression and response to treatment.

\section{Conclusion}
In our study with GDS3837 gene expression dataset, XGBoost demonstrated strong predictive performance, with an AUC score of 0.835, effectively identifying the most significant biomarkers associated with the condition under study. This outcome underscores the model's capability in handling complex genomic data and selecting relevant features, which are crucial for advancing our understanding of the underlying biological mechanisms.

\section{Discussion}
The biomarkers identified by the XGBoost model in our study, including CCAAT enhancer binding protein alpha (C/EBP$\alpha$), lactate dehydrogenase A4 (LDHA), UNC-45 myosin chaperone B (UNC-45B), checkpoint kinase 1 (CHK1), and hypoxia-inducible factor 1 subunit alpha (HIF-1$\alpha$), have been independently validated in the literature as being significantly associated with lung cancer. 

CCAAT/enhancer-binding protein alpha (C/EBP$\alpha$) is a transcription factor that plays a crucial role in regulating cellular differentiation and proliferation. Research has demonstrated that C/EBP$\alpha$ acts as a tumor suppressor in lung cancer, where its expression is often down-regulated. Studies using mouse models have shown that the deletion of C/EBP$\alpha$ significantly increases lung tumor incidence, particularly when induced by carcinogens like urethane. The reintroduction of C/EBP$\alpha$ expression in lung cancer cells has been associated with growth arrest and apoptosis, suggesting its potential as a therapeutic target in lung cancer treatment \cite{Zhang2013CEBPa}.

Lactate dehydrogenase A4 (LDHA) plays a crucial role in glycolysis, where cancer cells rely on glycolysis over oxidative phosphorylation even in the presence of oxygen. Elevated LDHA levels are associated with increased tumor growth and metastasis, contributing to the aggressive behavior of lung tumors by meeting the high metabolic demands of proliferating cancer cells. Similarly, UNC-45 myosin chaperone B (UNC-45B) and checkpoint kinase 1 (CHK1) are key proteins in lung cancer progression. Overexpression of UNC-45B enhances cellular motility and metastasis, while CHK1 supports cancer cell survival by repairing DNA damage. Targeting CHK1, particularly with DNA-damaging agents, and HIF-1$\alpha$, a factor promoting tumor growth in hypoxic conditions, presents promising therapeutic avenues in lung cancer treatment \cite{Fan2011LDHA, Guo2017UNC45B, Zhao2017CHK1, Semenza2012HIF1a}.

\bibliography{main}
\bibliographystyle{IEEEtran}

\end{document}